\documentclass[aapm,graphicx,groupedaddress,superscriptaddress]{revtex4-1}
\draft
\setcitestyle{super}
\usepackage[sort&compress]{natbib}
\usepackage{amsfonts}
\usepackage{amssymb}
\usepackage{amsmath,epsfig}
\usepackage{subfigure}
\usepackage{fmtcount}
\usepackage{array}
\usepackage{rotating}
\usepackage{fmtcount}
\usepackage{mathbbol}
\usepackage{dsfont}
\usepackage{placeins}
\usepackage{booktabs,xcolor,siunitx}
\usepackage{slashbox}
\usepackage{url}
\usepackage[caption=false]{subfig}
\usepackage{textcomp}
\usepackage{wrapfig}
\usepackage[letterpaper, portrait, margin=0.5in]{geometry}
\usepackage[mathlines]{lineno}
\usepackage{multirow}

\usepackage{threeparttable}
\usepackage{nomencl}
\usepackage{soul}
\usepackage{color}
\usepackage{algorithm}
\usepackage{algorithmic}
\usepackage{subfigure}
\usepackage{setspace}
\usepackage{algorithm}
\usepackage{algorithmic}

\makenomenclature

\usepackage{makeidx}
\makeindex 

\begin{document}
\def\vector{\underline}
\def\matrix{}
\renewcommand{\vec}[1]{\mathbf{#1}}
\newcommand{\mat}[1]{\mathbf{#1}}

\title{PEAR: PEriodic And fixed Rank separation for fast fMRI}

\author{Lior Weizman\footnote{Corresponding author, e-mail:weizmanl@tx.technion.ac.il, phone:+972-4-8291724}}
\affiliation{Department of Electrical Engineering, Technion - Israel Institue of Technology, Israel}
\affiliation{FMRIB Centre, University of Oxford, Oxford, United Kingdom}
\author{Karla L. Miller}
\affiliation{FMRIB Centre, University of Oxford, Oxford, United Kingdom}
\author{Yonina C. Eldar}
\address{Department of Electrical Engineering, Technion - Israel Institue of Technology, Israel}
\author{Mark Chiew}
\affiliation{FMRIB Centre, University of Oxford, Oxford, United Kingdom}

\date{\today}

\begin{abstract}
{\bf Purpose:}  In functional MRI (fMRI), faster acquisition via undersampling of data can improve the spatial-temporal resolution trade-off and increase statistical robustness through increased degrees-of-freedom.  High quality reconstruction of fMRI data from undersampled measurements requires proper modeling of the data. 
We present an fMRI reconstruction approach based on modeling the fMRI signal as a sum of periodic and fixed rank components, for improved reconstruction from undersampled measurements. 

{\bf Methods:} The proposed approach decomposes the fMRI signal into a component which a has fixed rank and a component consisting of a sum of periodic signals which is sparse in the temporal Fourier domain. Data reconstruction is performed by solving a constrained problem that enforces a fixed, moderate rank on one of the components, and a limited number of temporal frequencies on the other.  Our approach is coined PEAR - PEriodic And fixed Rank separation for fast fMRI.

{\bf Results:} Experimental results include purely synthetic simulation, a simulation with real timecourses and retrospective undersampling of a real fMRI dataset. Evaluation was performed both quantitatively and visually versus ground truth, comparing PEAR to two additional recent methods for fMRI reconstruction from undersampled measurements. Results demonstrate PEAR's improvement in estimating the timecourses and activation maps versus the methods compared against at acceleration ratios of R=8,16 (for simulated data) and R=6.66,10 (for real data).

{\bf Conclusions:}
This paper presents PEAR, an undersampled fMRI reconstruction approach based on decomposing the fMRI signal to periodic and fixed rank components.  PEAR results in reconstruction with higher fidelity than when using a fixed-rank based model or a conventional Low-rank+Sparse algorithm. We have shown that splitting the functional information between the components leads to better modeling of fMRI, over state-of-the-art methods.
\end{abstract}



\maketitle

\newpage

\section{Introduction}
\label{sec:intro}

Accelerating acquisition in functional MRI (fMRI) has gained significant attention in neuroimaging. Accelerated fMRI can provide data at higher frame rates or sampling bandwidths, leading to higher temporal degrees of freedom \cite{feinberg2010multiplexed}. This enables the use of more powerful and sophisticated analysis techniques \cite{smith2012temporally}. Alternatively, accelerated fMRI may be used to increase spatial resolution without sacrificing temporal fidelity, enabling time-resolved studies of the functional organization of the brain at
finer scales, like cortical layers \cite{koopmans2011multi} or columns \cite{yacoub2008high}. In resting state fMRI, where the goal is to estimate brain connectivity networks of the subject, accelerated data acquisition can improve the estimation of resting state networks (RSNs) \cite{beckmann2005investigations}. 

Numerous methods for accelerating acquisition of MRI data by exploiting its intrinsic structure and redundancy have been published. In clinical dynamic MRI (e.g. Cardiac MRI and Dynamic Contrast Enhanced (DCE) MRI), many methods are based on undersampling in the k-t space \cite{madore1999unaliasing,pruessmann1999sense,tsao2003k}. Since the introduction of compressed sensing (CS) \cite{eldar2012compressed,lustig2007sparse}, accelerated CS-based methods for clinical dynamic MRI have also been developed \cite{gamper2008compressed,jung2009k}.  While some of these methods have been adapted to fMRI \cite{jung2009performance,zong2014compressed,jeromin2012optimal,holland2013compressed,chavarrias2015exploitation,fang2015high,zhao2010low}, the different nature of the fMRI signal (when compared to clinical dynamic MRI), e.g. its low variance of signal of interest and its limited spatial compressibility, limits the adoption of those implementations. 

Recently, we introduced an approach for reconstruction of fMRI from undersampled measurements that is based on modeling fMRI as fixed-rank, i.e., k-t FASTER (FMRI Accelerated in Space-time by means of Truncation of Effective Rank) \cite{chiew2015k}. It aligns with the common analysis approaches in fMRI, which estimate limited numbers of spatial and temporal components from the data. It has been demonstrated that k-t FASTER provides higher quality of activation and resting state maps when compared with other methods for fMRI reconstruction from undersampled data. This method has been extended recently to include multi-channel coil sensitivity information and more flexible radial-Cartesian sampling \cite{chiew2016accelerating}, providing additional encoding information and more incoherent sampling of k-space, resulting in robust recovery of task-based fMRI data at acceleration factors of $10$ times or higher.

Several other approaches for reconstruction of fMRI from undersampled measurements have been recently suggested~\cite{aggarwal2016accelerated,singh2015under,petrov2016improving,aggarwal2017optshrink,lam2013accelerated}. Aggarwal et al. examined enforcing a low-rank  model  and signal sparsity \cite{aggarwal2016accelerated}. Others explored exploiting low rank and sparsity after the separation of fMRI into two components. This approach, known as ”Low-rank plus Sparse” (or L+S) \cite{candes2011robust}, consists of modeling the data as a sum of two components, where low rank is enforced on one of them, and sparsity in some transform domain is enforced on the other. L+S has been examined for both clinical dynamic MRI \cite{otazo2015low} and fMRI \cite{singh2015under,otazo2015Lowrankplus,petrov2016improving,aggarwal2017optshrink}.  




The common implementation of L+S for clinical dynamic MRI and fMRI consists of modeling the low rank component as a ``background" image while the sparse component contains the dynamic information. This approach leads to satisfactory results for clinical dynamic MRI applications where the dynamic signal is significantly above the noise level (e.g. MR angiography (MRA) and DCE-MRI) or periodic in the time domain (e.g. cardiac MRI). However, based on our experiments, its performance for fMRI, where the signal is often near the noise level and filtered by the hemodynamic response, is sub-optimal. 


In  this  study,  we  examine a different separation of the data, where both components contain functional information.
While most previous methods that combine low rank and sparsity are based on solving an unconstrained minimization problem by singular value soft-thresholding (SVT) \cite{cai2010singular}, in our approach we solve a constrained  minimization problem based on truncating the singular values (a.k.a Truncated SVD or TSVD) \cite{gavish2014optimal}. Our approach forces  one of the components to have a moderate fixed rank, and the other to be sparse in the temporal Fourier domain, leading to improved results compared to an SVT-based approach. Reconstruction is performed  via alternating minimization, that enforces the fixed-rank requirement and sparsity iteratively. 



We call our approach PEAR: PEriodic And fixed Rank separation for fast fMRI. We examine reconstructions from undersampled data acquired using golden-angle radial sampling\cite{winkelmann2007optimal}, and correspondence to both  time-course information (using General Linear modeling (GLM)) and spatial information (resting state network maps estimated via dual regression \cite{beckman2009dual}). Our experiments consist of a purely synthetic simulation, to show the concept of separation between the components, a synthetic simulation using real timecourses to examine correspondence of results to real and known time-courses, and retrospective sampling of a real resting state fMRI dataset, to examine resting state network recovery. We compare PEAR to k-t FASTER that uses a fixed-rank model only, and to a conventional, SVT-based L+S implementation. We also explore the contributions of the different components in PEAR. Based on our experiments, PEAR exhibits better estimation of the timecourses and resting state networks from undersampled data, compared to k-t FASTER and to L+S, using only 6.25\% of the data in the simulations and 10\% of the data in the retrospective sampling experiment. 

The paper is organized as follows. Section \ref{sec:method} presents the proposed method for faster fMRI via separation of signals into periodic and fixed rank components. Section \ref{sec:results} describes experimental results. Section \ref{sec:discussion} discusses theoretical aspects and implementation details of our method and Section \ref{sec:conclusions} concludes by highlighting the key results.

\section{Method}
\label{sec:method}

In MRI, data is acquired in the spatial Fourier domain (k-space). In dynamic MRI applications, such as  cardiac MRI, MRA and DCE MRI, as well as in fMRI, the k-space of each temporal frame is acquired. By undersampling k-space (i.e. taking only partial k-space measurements for each temporal frame), one can obtain higher frame rate, or alternatively, cover a greater extent in k-space, thereby increasing spatial resolution without decreasing temporal resolution.

In the problem of fMRI reconstruction from undersampled k-space, our goal is to recover the time series of acquired images. For simplicity, the time series is represented as a space-time matrix, $\mat{X}\in \mathbb{R}^{N \times T}$ where each column is a 3D temporal frame concatenated as a vector, $N$ denotes the number of pixels in a single frame, and $T$ denotes the number of frames in the time series. The measurement model, which takes into account that in most cases data is acquired using multiple coils is:
\begin{equation}
\vec{y}=\mat{E}\{\vec{X}\}+\mat{z}
\end{equation}
\noindent where $\vec{y}$ is a vector of undersampled measurements and $\vec{E}$ is a general linear operator that maps a matrix to a vector. For acquisition with multiple receiver coils, $\vec{E}$ consists of multiplication by coil sensitivities followed by an undersampled Fourier transform. The vector $\vec{z}$ represents the measurement noise, modeled as complex Gaussian with zero mean.

Since $\vec{y}$ is generated via undersampling, proper reconstruction of $\vec{X}$ from $\vec{y}$ requires assumptions on $\vec{X}$. Relying on framework of CS, many methods that are based on sparsity of $\vec{X}$ in some transform domain were examined for dynamic MRI in general and for fMRI in particular. In our recent work, we considered modeling $\mat{X}$ as a fixed rank matrix, which aligns with the theory that $\mat{X}$ is composed of a relatively small number of spatially coherent temporal processes. The fixed-rank based approach for fMRI (i.e., k-t FASTER) solves the following minimization problem \cite{chiew2015k}:
\begin{equation}
\begin{aligned}
& \underset{\vec{X}\in \mathbb{R}^{N \times T}}{\text{min}}
&  \|\vec{y}-\mat{E}\{\vec{X}\}\|_2 \ \ s.t. \ \  \text{rank}(\mat{X})=r,
\end{aligned}
\label{eq01}
\end{equation}
\noindent where $r$ is a fixed, moderate rank that ranges between 20 and 50 in our fMRI approach (but may be much lower in other MRI modalities). 
Unlike some other types of dynamic MRI that exhibit high variance of signals of interest, in fMRI valuable information may also be embedded in low variance components, slightly above the noise level. Consequently, method to retrieve information from higher dimensions is expected to provide better results for fMRI.

An approach that was applied initially for clinical dynamic MRI \cite{otazo2015low}, and has been examined recently for fMRI \cite{petrov2016improving,singh2015under,otazo2015Lowrankplus}, consists of modeling the dynamic sequence as a sum of two components. A low-rank component that represents mainly the background (denoted as the L component), and a component that contains the valuable signal. The latter is modeled as sparse in some transform domain, and denoted as the S component. This approach, known as L+S \cite{candes2011robust}, is based on solving the following unconstrained problem:

\begin{equation}
\begin{aligned}
& \underset{\vec{L},\vec{S}\in \mathbb{R}^{N \times T}}{\text{min}}
&  \frac{1}{2}\|\vec{y}-\mat{E}\{\vec{L}+\vec{S}\}\|_2^2+\lambda_1\|\vec{L}\|_*+\lambda_2\|\mat{\Psi}\{\vec{S}\}\|_1
\end{aligned}
\label{eq02}
\end{equation}
\noindent where $\mat{L}$ and $\mat{S}$ denote the low-rank and sparse components, $\|\cdot\|_*$ and $\|\cdot\|_1$ are the nuclear norm and the $\ell_1$ norm, $\lambda_{1,2}$ are tuning parameters that control the weight given to each term in the optimization problem and the reconstructed space-time matrix is $\mat{X}=\mat{L}+\mat{S}$. The linear transformation $\mat{\Psi}$ is a sparsifying transformation applied on $\mat{S}$, depending on the specific dynamic MRI application. For MRA, $\mat{\Psi}$ may be chosen as an identity transform, whereas for cardiac MRI, which consists of periodic temporal structure, $\mat{\Psi}$ may be a temporal Fourier transform (i.e., applying a Fourier transform row-wise, independently on each of the rows of $\mat{S}$) \cite{otazo2015low}. For fMRI, both types of transformations were examined: Otazo et el. \cite{otazo2015Lowrankplus} considered the identity transform (although their decomposition is used for analysis rather than acceleration) and Singh et al. \cite{singh2015under} examined the temporal Fourier transform.   We note that L+S solves an unconstrained problem that does not explicitly enforce a fixed rank, and the solution is often based on SVT. By viewing the results of current implementations of L+S for fMRI\cite{otazo2015Lowrankplus,singh2015under} we found that in practice the resulting L component tends to have a very low rank and typically contains only background information, while the important functional information is in the S component.

\begin{figure}
\includegraphics[width=460pt, trim={2cm 1cm 2cm 1cm},clip=true]{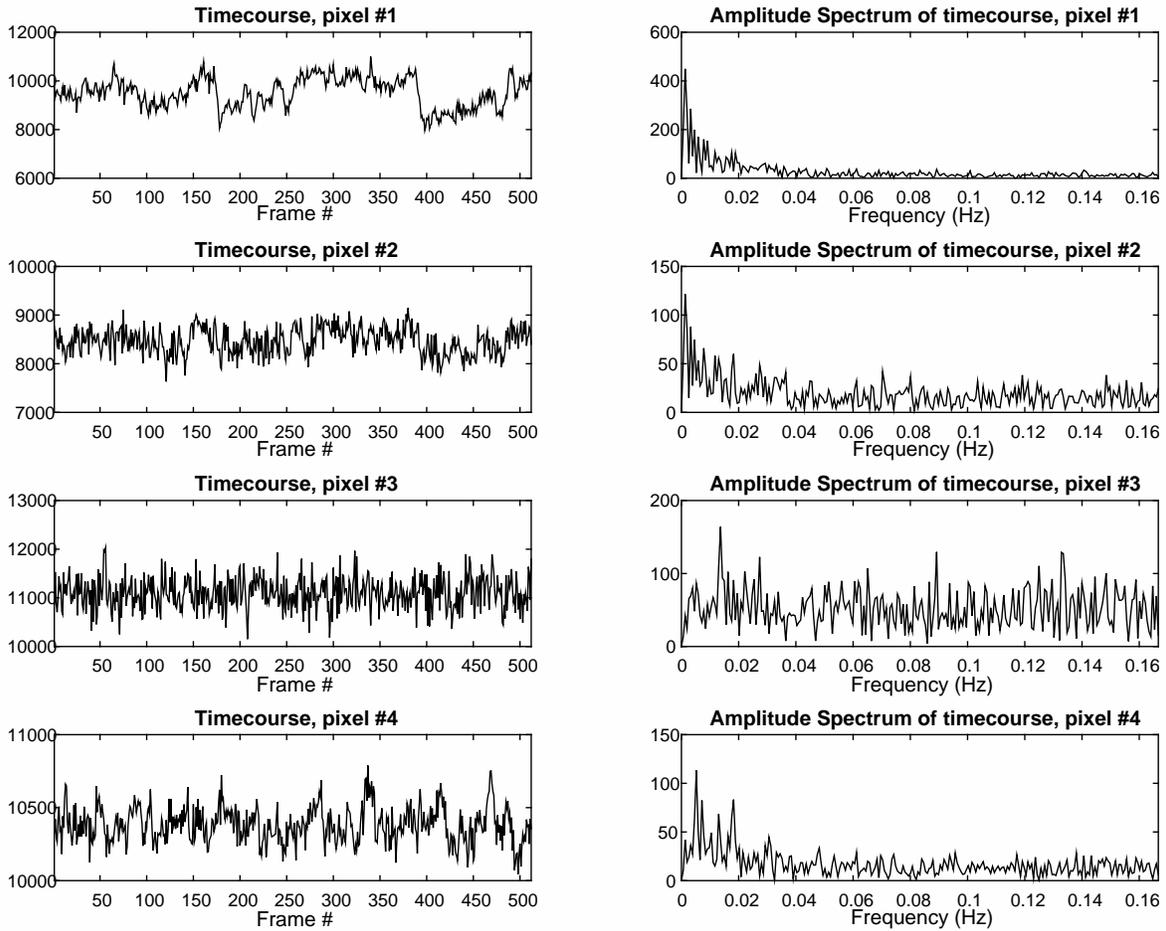}\\
\begin{minipage}[c]{0.3\textwidth}
\includegraphics[width=125pt]{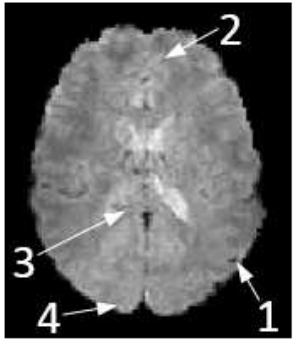}
  \end{minipage}
\begin{minipage}[c]{0.65\textwidth}
\caption{Top: Timecourses (left) and amplitude spectrum (right) after removing DC of four selected pixels from a preprocessed resting-state fMRI dataset. Bottom left: spatial locations of selected pixels. Pixels \#1-\#3 showed high correspondence ($|Z|>6$) with Default Mode Network and pixel \#4 showed high correspondence with a visual network. The amplitude spectrum of the pixels shows that while pixels \#1, \#2 and \#4 exhibit a limited number of peaks in the spectral domain (and therefore may be suitable for sparse modelling in the temporal Fourier domain) pixel \#3 involves a wide range of spectral components. As a result, separation into a fixed-rank and periodic components as proposed by our method would allow better representation of those signals, compared to using a fixed rank component (k-t FASTER) or a periodic component only (L+S). 
}
\label{fig1}
  \end{minipage}
\end{figure}

Some fMRI analysis models suggest that while neural signals have strong band limited components, they also exist across the frequency spectrum \cite{thompson2014quasi}. Therefore, we consider including sparsity in the temporal spectrum, to capture the bandlimited assumption, in  addition to a fixed rank representation that would be more suitable for broader-band signals. In particular, we propose modeling the fMRI signal as a sum of a fixed rank component, which contains the high variance information, and a periodic component that captures the periodicity that is not captured in the fixed rank component. 
Thus, we model the fMRI data $\mat{X}$ as $\mat{X}=\mat{A}+\mat{P}$, where $\mat{A}$ and $\mat{P}$ are the fixed rank and periodic components, respectively. 
We enforce a limited number of temporal periodic signals for $\mat{P}$, by demanding sparsity in the temporal Fourier domain, and a fixed rank for $\mat{A}$ which contains the high variance signal.

To understand the rationale behind this modeling, Fig. \ref{fig1} shows 4 timecourses of arbitrarily selected pixels in a resting state fMRI dataset, that exhibited high correspondence ($|Z|>6$) with regressors that represent a Default Mode Network (DMN) map or a visual network map after a dual-regression against those network maps. The signals in the figure were extracted from the fMRI sequence after conventional fMRI pre-processing (including skull stripping, motion correction and slice timing correction), and the spatial locations of the selected pixels are shown at the bottom left of the figure. The timecourses are shown in the time domain (top left) and in the temporal Fourier domain, after removing the DC component (top right). It can be seen that while three of the timecourses exhibit  peaks in the spectral domain and are suitable for sparse representation in the temporal Fourier domain, one of the timecourses exhibits a broad spectrum in the temporal Fourier domain. As a result,  separation into fixed rank and periodic components allows better representation of signals (compared to fixed-rank only or periodic component only) as it is expected to capture both broad-band and band-limited temporal spectra. 

To obtain the separation into A and P components, we propose the following minimization problem:
\begin{equation}
 \underset{\substack{\mat{A}\in \mathbb{C}\\ \quad \; \; \;    \mat{P}\in \mathbb{R}^{N \times T}}}{\text{min}} \frac{1}{2}\|\vec{y}-\mat{E}\{\vec{A}+\vec{P}\}\|_2^2+\lambda\|\mat{F}_t\{\vec{P}\}\|_1
\label{eq04}
\end{equation}
\noindent where $\mathbb{C}$ is the set of matrices with a fixed rank $r$ (ranges between 20 and 50 in our fMRI approach) and $\mat{F}_t$ is the temporal Fourier transform that applies a Fourier transform row-wise on each of the rows of $\mat{P}$ (where $\hat{\mat{X}}=\hat{\mat{A}}+\hat{\mat{P}}$). 

We solve (\ref{eq04}) using alternating minimization (AM) \cite{niesen2009adaptive}. In this approach, in each iteration we perform minimization with respect to one variable while keeping the other one fixed, and then switch between the variables. In our case, we start with an arbitrary initial point $\mat{P}_0$. For $n \ge 1$ we iteratively compute:
\begin{equation}
\mat{A}_n=\underset{\mat{A}\in \mathbb{C}}{\text{arg min}} \ D(\mat{A},\mat{P}_{n-1})=\underset{\mat{A}\in \mathbb{C}}{\text{arg min}} \ \frac{1}{2}\|\vec{y}-\mat{E}\{\vec{A}+\vec{P}_{n-1}\}\|_2^2
\label{eq21}
\end{equation}

\begin{equation}
\mat{P}_n=\underset{ \mat{P}\in R^{N \times K}}{\text{arg min}} \ D(\mat{A}_n,\mat{P})=\underset{ \mat{P}\in R^{N \times K}}{\text{arg min}}  \frac{1}{2}\|\vec{y}-\mat{E}\{\vec{A}_n+\vec{P}\}\|_2^2+\lambda\|\mat{F}_t\{\vec{P}\}\|_1.
\label{eq22}
\end{equation}
\noindent We solve each sub-problem (\ref{eq21},\ref{eq22}) via gradient projection \cite{bertsekas1999nonlinear} where the proximal gradient is used for the non-differentiable $\ell_1$ function in (\ref{eq22}). A solution for (\ref{eq21}) has been proposed by Goldfarb et al. \cite{goldfarb2011convergence}, a.k.a Iterative Hard Thresholding with Matrix Shrinkage (IHT-MS). It consists of a gradient step for data consistency followed by a projection step onto the subspace $\mathbb{C}$. The general step is:
\begin{equation}
\mat{A}_n=R_r(S_{\mu}(\mat{A}_{n-1}-\alpha\mat{E}^{H}\{\mat{E}\{\mat{A}_{n-1}+\mat{P}_{n-1}\}-\mat{y}\}))
\label{eq20}
\end{equation}
where $R_r(\mat{Q})$ is the projection onto the subspace $\mathbb{C}$, defined as: $R_r(\mat{Q})=\sum_{i=1}^r \sigma_i \vec{u}_i \vec{v}_i^H$ where $\sigma_1\geq \sigma_2 \geq...\geq \sigma_m$ are the singular values of $\mat{Q}$, and $\mat{u}_i$ and $\mat{v}_i$ are the singular vectors associated with $\sigma_i$. The operator $S_\mu(\mat{Q})$ is the singular value soft-thresholding operator, defined as: $S_\mu(\mat{Q})=\mat{U}[\mat{\Sigma}-\mu\mat{I}]_{+}\mat{V}^H$ where $\mat{\Sigma}=\text{diag}(\sigma_1, \sigma_2,...,\sigma_m)$ and $\mat{U}$ and $\mat{V}$ are the left and right singular vectors associated with $\mat{\Sigma}$. The parameter $\alpha$ is a step size, which controls the rate of convergence.

The rationale behind applying $S_{\mu}(\cdot)$ before applying $R_r(\cdot)$ can be explained by modelling the lower singular values,  $\{\sigma_i\}_{i=r+1}^{m}$ as representing noise, while $\{\sigma_i\}_{i=1}^{r}$  represent data contaminated with additive noise.  The subtraction of $\mu$ from $\{\sigma_i\}_{i=1}^{r}$ is explained as removing noise from the higher singular values. Based on our experiments \cite{chiew2015k}, the selection of $\mu=c\cdot\sigma_{r+1}$ where $c$ is a fixed parameter, leads to good results. 

To solve ($\ref{eq22}$) we use the Iterative Shrinkage-Thresholding Algorithm (ISTA) \cite{daubechies2004iterative,beck2009fast}, whose details are given in Appendix \ref{sec:appendix}. The general step for the solution of (\ref{eq22}) is:
\begin{equation}
\mat{P}_{n}=\mat{F}_t^H\{\Lambda_\lambda(\mat{F}_t\{\mat{P}_{n-1}-\alpha\mat{E}^H\{\mat{E}\{\mat{A}_{n-1}+\mat{P}_{n-1}\}-\vec{y}\}    \})\}
\label{eq25}
\end{equation}
\noindent where $\Lambda_{\lambda}$ indicates the soft-thresholding operator with parameter $\lambda$, applied element-wise. Finally, by defining $\mat{X}_{n-1}=\mat{A}_{n-1}+\mat{P}_{n-1}-\alpha\mat{E}^H\{\mat{E}\{\mat{A}_{n-1}+\mat{P}_{n-1}\}-\vec{y}\}\}$ we get that (\ref{eq20}) and (\ref{eq25}) become:
\begin{equation}
\mat{A}_n=R_r(S_{\mu}(\mat{X}_{n-1}-\mat{P}_{n-1}))
\end{equation}
\begin{equation}
\mat{P}_n=\mat{F}_t^H\{\Lambda_\lambda(\mat{F}_t\{\mat{X}_{n-1}-\mat{A}_{n-1}\})\}.
\end{equation}
The convergence of an iterative solution based on (\ref{eq20}) has been proven by Goldfarb et al. and the convergence of an iterative solution based on (\ref{eq25}) is well studied in the literature \cite{daubechies2004iterative,combettes2005signal}. Therefore, based on Csiszar and Tusnády \cite{csisz1984information}, the convergence of our proposed AM framework is guaranteed. 

The proposed algorithm is coined PEriodic And fixed Rank separation for fast fMRI (PEAR) and is summarized in Algorithm 1, where $\text{SVD}$ represents the singular value decomposition and $\{\sigma_i\}_{i=1}^{r+1}$ are the singular values of the matrix $\mat{X}_{n-1}-\mat{P}_{n-1}$ in descending order. The operator $\mat{F}^H_t$ is the conjugate temporal Fourier transform, and $\mat{E}^H$ is the conjugate transpose of $\mat{E}$.

\newlength\myindent
\setlength\myindent{10em}
\newcommand\bindent{%
  \begingroup
  \setlength{\itemindent}{\myindent}
  \addtolength{\algorithmicindent}{\myindent}
}
\newcommand\eindent{\endgroup}
\renewcommand{\algorithmicrequire}{\textbf{Input:}}
\renewcommand{\algorithmicensure}{\textbf{Output:}}
\begin{algorithm}[H]
\caption{PEAR: PEriodic And fixed Rank separation for fast fMRI}
\label{algo1}
\begin{algorithmic}
\REQUIRE \hspace{3mm} \\
Multicoil undersampled k-t data: $\vec{y}$ \\
Space-time multicoil encoding operator: $\vec{E}$\\
Temporal Fourier transform: $\mat{F}_t$\\
Predefined rank: $r$, \ Soft-shrinkage parameter: $c$ \\
Tuning constant: $\lambda$, \ Step size: $\alpha$\\ 
Iteration limit: $N$
\ENSURE Estimated fMRI time-series: $\mat{\hat{X}}$
\renewcommand{\algorithmicrequire}{\textbf{Initialize:}}
 \REQUIRE \hspace{3mm} \\
\STATE $\vec{P}_0=0$, $\vec{X}_0=\vec{E}^H\{\vec{y}\}$ 
\renewcommand{\algorithmicrequire}{\textbf{Iterations for} $n=1..N$}
 \REQUIRE \hspace{3mm} \\
\STATE    $\vec{U}\vec{\Sigma}\vec{V}^H=\text{SVD}(\mat{X}_{n-1}-\mat{P}_{n-1}$)
\STATE     $\vec{\Sigma}(j,j)=\begin{cases}
    \vec{\Sigma}(j,j)-c\cdot\sigma_{r+1}, & j<r \ \text{and} \ \vec{\Sigma}(j,j)>c\cdot\sigma_{r+1}\\
  0, & \text{otherwise}.
  \end{cases}$
\STATE    $\vec{A}_n=\vec{U}\vec{\Sigma}\vec{V}^H$
\STATE $\mat{P}_n=\mat{F}_t^H\{\Lambda_\lambda(\mat{F}_t\{\mat{X}_{n-1}-\mat{A}_{n-1}\})\}$
\STATE $\mat{X}_n=\mat{A}_{n}+\mat{P}_{n}-\alpha\mat{E}^H\{\mat{E}\{\mat{A}_{n}+\mat{P}_{n}\}-\vec{y}\}\}$
\end{algorithmic}
\end{algorithm}
To summarize, the major differences between L+S and the PEAR approach for fMRI are outlined below:
\begin{itemize}
\item {\bf Thresholding mechanism and a fixed rank solution:} The solution of the L+S problem given in (\ref{eq02}) using the same AM approach used for solving (\ref{eq04}) results in an algorithm that is different from Algorithm 1 in the thresholding mechanism of the singular values. While in Algorithm 1 we perform singular value soft-tresholding (SVT) with the value $c\cdot \sigma_{r+1}$ followed by truncating the $r+1...m$ singular values (TSVD), for the solution of (\ref{eq02}) we perform SVT with value $\lambda_1$, with no rank constraint. The issue of an SVT-based solution versus a TSVD-based solution has been studied in the literature previously \cite{josse2013selecting}, and it has been shown that TSVD performs better for fixed rank problems \cite{josse2013selecting}. Based on our experience, fMRI can be considered as a fixed rank problem, since the number activation networks is relatively small and in many cases kept fixed for analysis.  This statement also aligns with our experimental results for fMRI hereinafter.   
\item {\bf Separation of functional information between components:} In conventional implementation of L+S for fMRI \cite{singh2015under,petrov2016improving,otazo2015Lowrankplus}, the L component is modeled as very low rank, and therefore contains background information and no functional information. In PEAR, functional information is split between the components. Consequently, recovered functional information is not limited to periodic signals only, and results are improved compared to L+S, as will be shown in the next section. 
Indeed, the solution of (\ref{eq04}) can be approximated by solving (\ref{eq02}) with appropriate selection of  $\lambda_{1,2}$ values. However, PEAR enables enforcing a fixed rank (based on a priori knowledge of typical fMRI dimensionalities, often between 20-50) for a variety of datasets, obviating the need to examine a range of $\lambda_{1,2}$ values for each separate dataset. In addition, our experiments show that solving (\ref{eq04}) provides better results for fMRI, when compared to solving (\ref{eq02}) also for the case where $\lambda_{1,2}$ were carefully chosen for optimality.

\end{itemize}


\section{Experimental Results}
\label{sec:results}

To demonstrate the performance of PEAR compared to a well defined ground truth and additional algorithms for the reconstruction of fMRI from undersmpled measurements, we performed 3 types of experiments. In all experiments PEAR is compared to k-t FASTER \cite{chiew2015k} and L+S \cite{singh2015under} (where sparsity is enforced in the temporal Fourier domain, as shown in (\ref{eq02})).

The first experiment is a simulation based on synthetically generated mixtures of periodic and aperiodic signals, and aims to compare the results of PEAR to the aforementioned methods and to examine how PEAR separates the signals into periodic and fixed rank components. The second experiment is an extension of the first experiment using realistic time-courses instead of purely synthetic ones. In the third experiment we examine the performance of PEAR for a retrospectively undersampled realistic 3D fMRI data sequence.

In all experiments, data is undersampled retrospectively, through a radial sampling approach using the golden-angle \cite{winkelmann2007optimal,graedel2016motion} and the NUFFT \cite{fessler2003nonuniform} package was used for forward and adjoint spatial Fourier transforms. The output of each experiment is provided as z-statistics maps that reflect the degree to which each timecourse (in the case of experiments 1 and 2) or spatial regressor (in case of experiment 3) is expressed with a unique time-course in the data. Output maps were null-corrected using a Gaussian and Gamma mixture model \cite{beckmann2005investigations}.

The parameters $c=0.7$, $\alpha=1$ (for k-t FASTER) and $\alpha=0.5$ (for PEAR and L+S) were selected experimentally. In all cases, all time-points were initialized to the mean image calculated from all projections. All algorithms were run 100 iterations or until the minimum update between consecutive iterations was below $10^{-4}$. For k-t FASTER and PEAR we examined the parameter $r$ in the range between 1 and 50, and experimentally used $r=32$ for k-t FASTER, $r=27$ for PEAR in experiments 1 and 2, and $r=20$ for PEAR in experiment 3. In experiments 1 and 2 the parameters were tuned for optimal performance for each method (where optimal performance is evaluated by examining the z-stat maps), and in experiment 3 parameters were tuned for optimality on a training set and results were obtained using the same parameters for an unseen fMRI sequence.

\subsection{Experiment 1: Purely synthetic simulation}
\begin{figure}
\includegraphics[width=490pt]{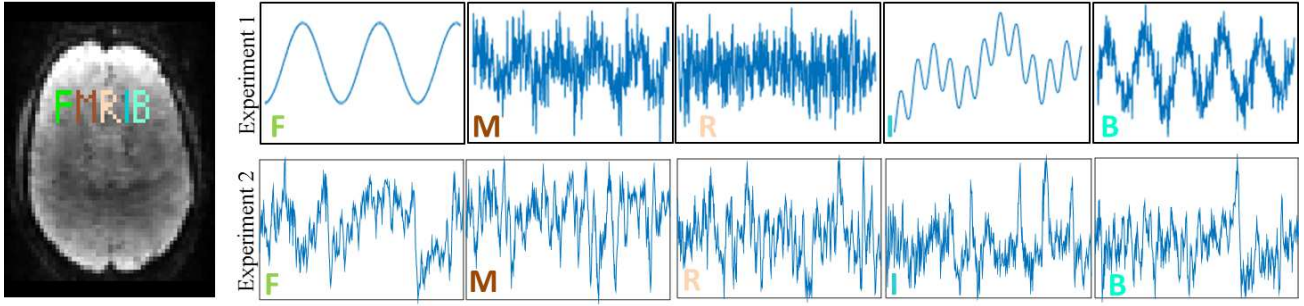}\\
\caption{Left: Spatial locations of the ROIs used in experiments 1 and 2, each ROI is formed as single letter and contains a single timecourse. Top: The timecourses used for each ROI in experiment 1. ``F" and ``I" are purely periodic timecourses where ``F" contains a single frequency and ``I" contains a mixture of three frequencies, ``R" is a purely aperdioc timecourse, and ``M" and ``B" are superposition of periodic and aperiodic timecourses. Bottom: Timecourses used in experiment 2 for each simulated ROI (letter).}
\label{fig2}
\end{figure}

In this experiment, we simulated a phantom consisting of 5 Regions of Interest (ROIs). Each ROI is spatially formed as a single letter from the letters ``FMRIB", and contains one of 5 purely synthetic timecourses, generated as follows. The letters ``F" and ``I" were purely periodic timecourses where ``F" contains a single frequency and ``I" contains a mixture of three frequencies, ``R" was a purely aperdioc timecourse, and ``M" and ``B" were a superposition of periodic and aperiodic timecourses. The phantom was added to a realistic background fMRI dataset, to form a 2D fMRI sequence with known functional timecourses, of size $64\times 64$, with $512$ time points. The timecourses and their spatial locations in the simulated image are shown in Fig. \ref{fig2} (left and top).  

Undersampling was carried out in the k-t space. We examined two undersampling ratios, first by taking 8 radial projections at each timepoint (corresponding to acceleration ratio of R=8 relative to a fully-sampled, maximally efficient Cartesian acquisition). We then repeated the experiment using only 4 radial projections at each timepoint (corresponding to R=16). This simulates one slice of a hybrid radial-Cartesian trajectory, which rotates an EPI trajectory within 3D k-space \cite{chiew2016accelerating}. An additive white Gaussian noise with zero mean was added to the samples in the k-space domain to obtain SNR of $25$dB. 
For PEAR, $\lambda$ was examined in the range of 0.45-3.4 and was selected as $\lambda=0.91$ experimentally. For L+S, $\lambda_{1}$ was examined in the range of 1.1-3.4 and $\lambda_{2}$ was examined in the range of 0.45-3.4. These parameters were selected as $\lambda_1=1.6$ and $\lambda_2=0.91$ experimentally (the values for $\lambda,\lambda_{1,2}$ are provided after normalization with respect to the standard deviation of the data).

\begin{figure}
\includegraphics[width=520pt]{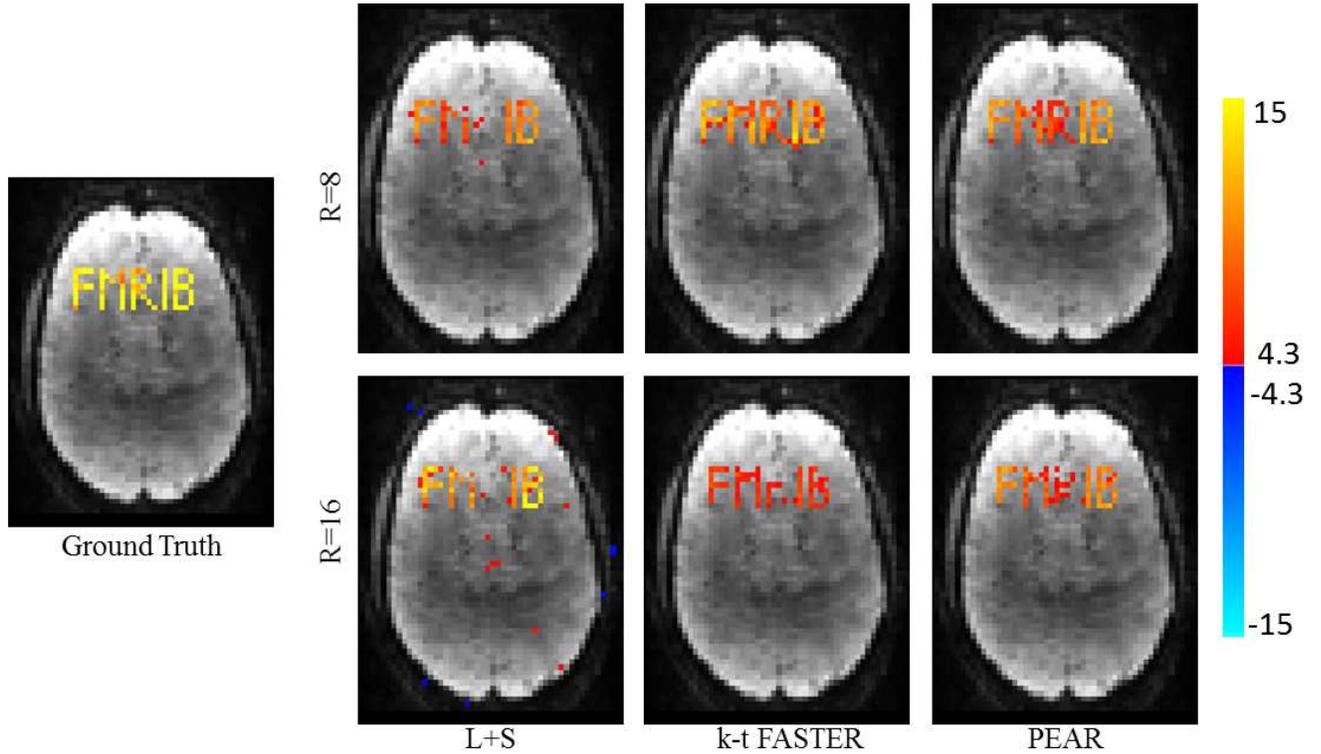}
\caption{Experiment 1: GLM F-test results of L+S, k-t FASTER and PEAR for purely synthetic simulation, for R=8 (top) and R=16 (bottom). The z-stat map of the ground truth is also shown (left). All maps are thresholded at $|Z|>4.3$ and with color scale mapped between $4.3<|Z|<15$. It can be seen that L+S is unable to recover the letter ``R", due to its purely aperiodicity. While PEAR exhibits better results for R=8 and R=16 when compared to the other methods, the difference between k-t FASTER and PEAR is emphasized for R=16, where PEAR provides the most reliable result, with almost perfect recovery of the letters ``M" and ``B", at minimal ratio of false positive errors. }
\label{fig3}
\end{figure}

To examine the correspondence of the reconstruction with the ground-truth, we performed regression against the original timecourses using General Linear Model (GLM)\cite{jenkinson2012fsl}. Figure \ref{fig3} shows the F-test  results as null-corrected z-statistics maps\cite{beckmann2005investigations}, for the ground-truth data (fully sampled image without the addition of noise), L+S, k-t FASTER and PEAR, for both R=8 and R=16. All maps are thresholded at $|Z|>4.3$ and shown with color scale mapped between $4.3<|Z|<15$. 

It can be seen that PEAR provides the most reliable result, being the only method that almost perfectly recovers both ``M" and ``B" with minimum false positive errors at R=16. In addition, we see that L+S is unable to recover the aperiodic timecourse ``R", as opposed to both k-t FASTER and PEAR thanks to their fixed-rank component.

\begin{figure}
\begin{minipage}[l]{0.5\textwidth}
\includegraphics[width=230pt]{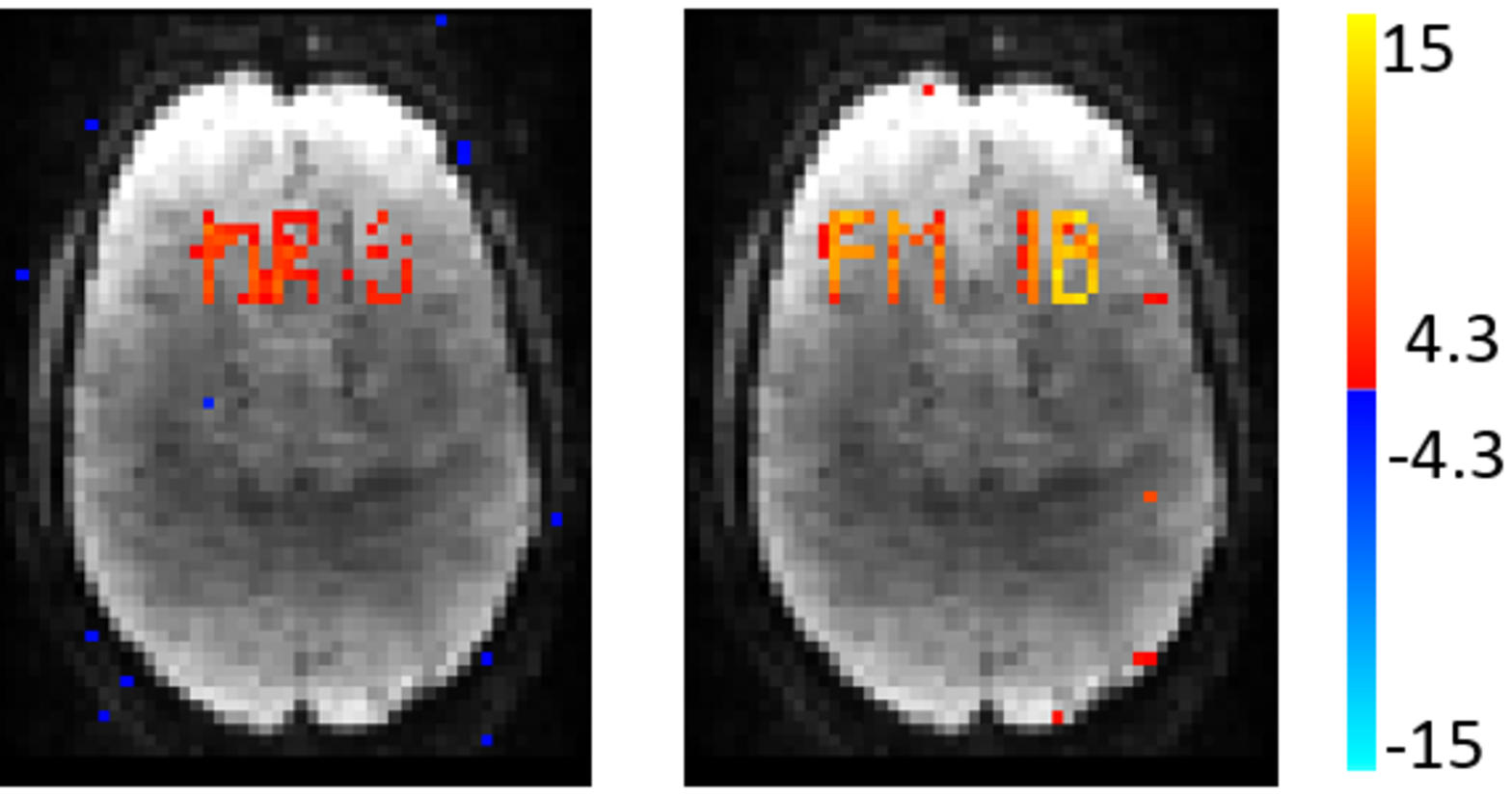}
\begin{flushleft}
\vspace{-5mm}
\hspace{0mm} PEAR: A component \hspace{1mm} PEAR: P component
\end{flushleft}
\end{minipage}
\begin{minipage}[l]{0.45\textwidth}
\caption{Experiment 1: GLM F-test results of A and P components of PEAR for R=16. Maps are thresholded at $|Z|>4.3$ and with color scale mapped between $4.3<|Z|<15$. It can be seen that the A component captures the letter ``R" which is purely aperiodic, and the letters ``M" and ``B" that include an aperiodic part. The P component captures the letters ``F" and ``I" which represent periodic timecourses, and the letters ``M" and ``B" that include an periodic part.
}\label{fig4}
\end{minipage}
\end{figure}

\begin{figure}
\includegraphics[width=500pt, trim={6.1cm 0cm 4.1cm 0cm},clip=true]{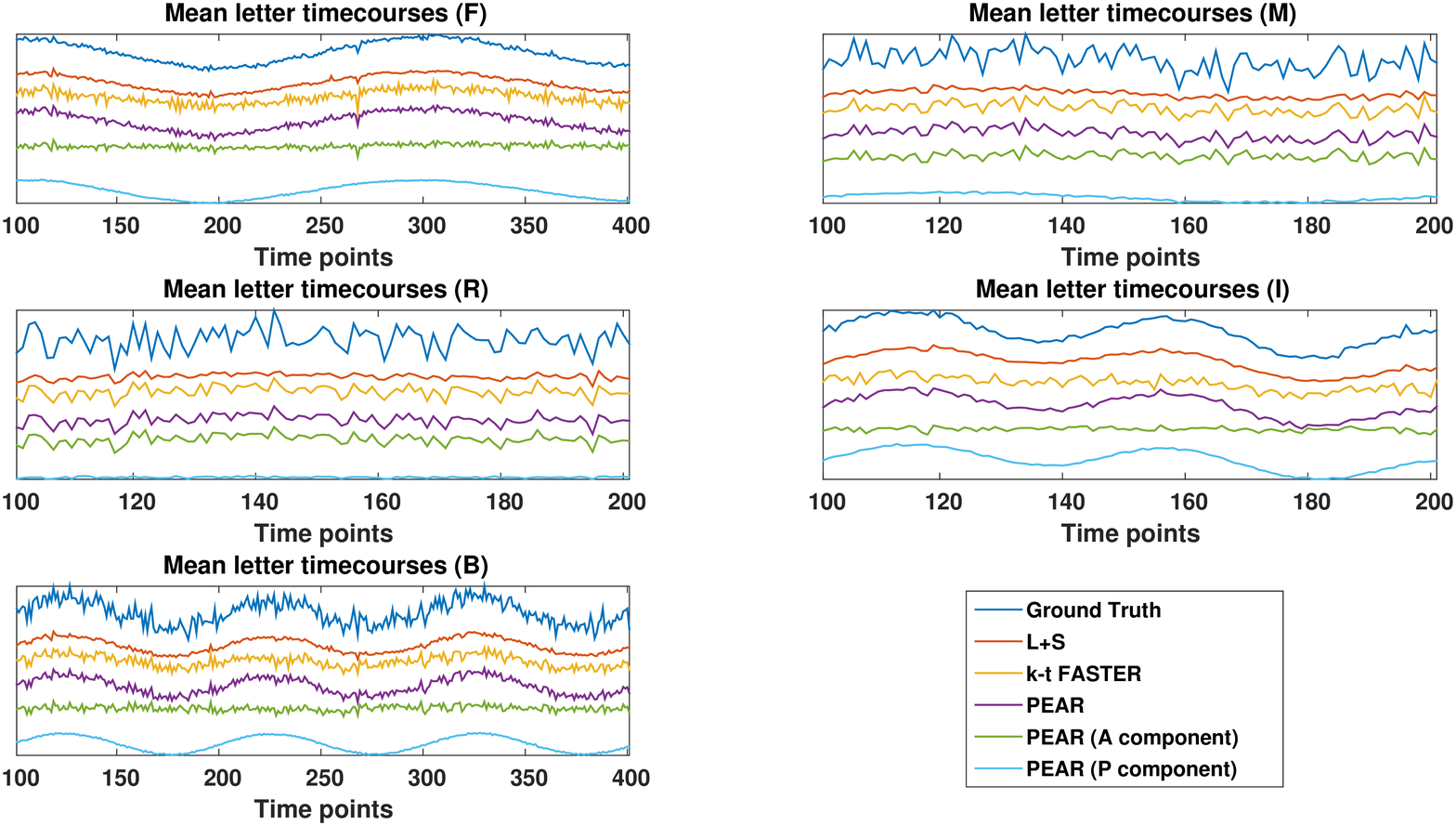}
\caption{Example portions of the mean timecourses of the five letter ROIs are shown for the ground truth, L+S, k-t FASTER and PEAR reconstruction results, including the timecourses for the A and P components of PEAR separately (Experiment 1, R=16). The timecourses are shown in arbitrary units, to allow proper examination of their structure. It can be seen that L+S is limited in its ability to track the rapid changes that appear in the letter ``R". In addition, the separation of PEAR into A and P component is clearly seen, as the P component exhibits high correspondence with letters that are fully periodic (``F" and ``I"), and A exhibits high correspondence with the aperiodic letter, ``R".}
\label{fig4a}
\end{figure}

An interesting analysis is the contribution of each component in PEAR. Figure \ref{fig4} shows the GLM results for the A and P components of PEAR separately (for R=16), where the z-statistics maps are thresholded at $|Z|>4.3$ and shown with color scale mapped between $4.3<|Z|<15$. Note that the sum of the null-corrected z-statistics maps of A and P is not equal to the z-statistics map of PEAR, due to the null-correction applied for each map, that depends with each map's noise level. However, the separation of PEAR into periodic and fixed rank components is clearly demonstrated. The A component highly corresponds with the letter ``R" which is a purely aperiodic timecourse, and with the letters ``M" and ``B" that include an aperiodic part. The P component highly corresponds to the letters ``F" and ``I" which are purely periodic timecourses, and to the letters ``M" and ``B" that include an periodic part. As demonstrated, this separation allows better modelling and leads to better recovery compared to k-t FASTER and L+S. Another analysis is presented in Fig. \ref{fig4a}, where example portions of the mean timecourses from the five letter ROIs are shown for the ground truth, L+S, k-t FASTER and PEAR reconstruction results, including the timecourses for the A and P components of PEAR separately. The timecourses are shown in arbitrary units, to allow proper examination of their structure. It can be seen that as expected, L+S is limited in its ability to track the rapid changes that appear in the letter ``R". In addition, the P component of PEAR indeed contains the periodic part of the signal, and therefore exhibits high correspondence with letters that are fully periodic (``F" and ``I"). 

These simulations clearly demonstrate the expected behaviour of our proposed approach under conditions where signals include pure periodicity; however, these are not a realistic depiction of fMRI data, even in task conditions, since these signals are rarely strongly periodic. The following experiment examines the performance of the various algorithms for realistic timecourses. 


\subsection{Experiment 2: Simulation with realistic timecourses}
In this experiment, we generated 5 realistic timecourses. First, we used a regression of a realistic fMRI dataset taken from the Human Connectome Project  (HCP) database \cite{van2013wu} against 15 canonical resting state network maps (RSNs) \cite{beckman2009dual}, derived from high-dimensional group-level Independent Component Analysis (ICA) of resting state fMRI datasets. The regression result provided 15 timecourses, each one corresponds to a single RSN regressor. We pulled 5 timecourses and used them instead of the purely synthetic timecourses used in experiment 1. These timecourses were used for the same simulation of the letters FMRIB, rather than retained in the original network spatial maps, due to the ease of visually evaluating the output parameter maps.

We repeated the same setting of experiment 1 (including the same SNR, sampling ratios, and selected parameters for each algorithm) where the only difference is the use of realistic timecourses instead of simulated ones. The realistic timecourses, including the corresponding regressors used for generation of the timecourses, are shown in Fig.~\ref{fig2} (bottom and left). 

\begin{figure}
\includegraphics[width=530pt]{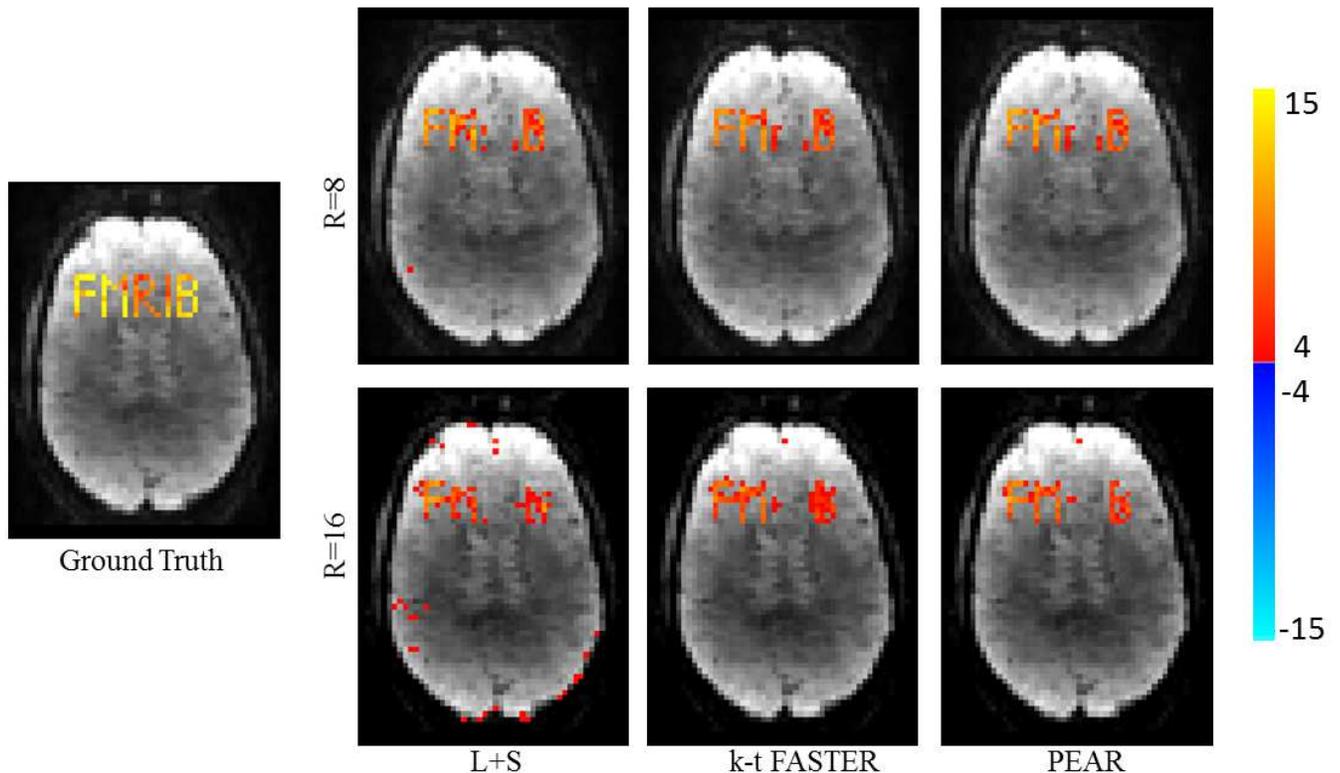}
\caption{Experiment 2: GLM F-test results of ground truth, L+S, k-t FASTER and PEAR for simulation with realistic timecoureses, for R=8 (top)and R=16 (bottom). The z-stat map of the ground truth is also shown (left). All maps are thresholded at $|Z|>4$ and with color scale mapped between $4<|Z|<15$. It can be seen that although ``R" and ``I" are almost irrecoverable and PEAR and k-t FASTER provide similar results for R=8, PEAR provides better results for R=16 with minimal ratio of false positive errors. }
\label{fig6}
\end{figure}

Figure \ref{fig6} shows the General Linear Model (GLM) F-test \cite{jenkinson2012fsl} results as null-corrected z-statistics maps that were computed against the realistic time courses of all letters, for the ground-truth data (fully sampled image without the addition of noise), L+S, k-t FASTER and PEAR, for R=8 and R=16. All maps are thresholded at $|Z|>4$ and shown with color scale mapped between $4<|Z|<15$. 

In correspondence with experiment 1, for realistic timecourses we see that for R=8, both k-t FASTER and PEAR provide similar results that outperform L+S. For R=16, we see that PEAR provides cleaner results, as can be seen mainly when comparing the recovery of the letters ``F" and ``B". It can also be seen that all methods are unable to recover ``R" and ``I" due to the high undersampling ratio. It can also be seen that ``R" and ``I" are the letters with the lowest Z values in the fully-sampled, ground truth data, due to their low energy in this experiment. 

\subsection{Experiment 3: Retrospective undersampling of real fMRI dataset}
In this experiment, we examined kt-FASTER, L+S and PEAR for the scenario of undersampled 3D fMRI data, taken from the HCP database  \cite{van2013wu}. Data was registered to an MNI standard space with dimensions $91 \times 109 \times 91$ and included 512 timepoints. We examined two sampling ratios. First, we used 15\% (R=6.66) of the data by taking only 14 radial blades at each timepoint for each axial slice. In addition, we examined the scenario of using only 10\% of the data (R=10). To keep memory requirements under control, reconstructions were performed independently for each 91x109 2D axial slice with 512 timepoints. After reconstruction, all 2D reconstructed slices were stacked together to form a 3D image with 512 time point for further analysis.  

After data reconstruction, we evaluated correspondence of the data to 15 canonical Resting State Networks (RSN) derived from high-dimensional group-level ICA of resting fMRI datasets from the HCP database  \cite{van2013wu}. Evaluation was done using dual regression \cite{beckmann2009group} as follows. First, we performed spatial regression of the reconstructed dataset against the canonical maps (regressors) to extract the timecourses corresponding to each map. Then, we performed temporal regression of the dataset against the time series. The output is a set of z-statistic maps (one for each regressor) that reflect the degree to which each spatial regressor is expressed with a unique time-course in the data. 

We compared the z-statistics maps from PEAR to those computed from the fully sampled data (ground truth) and from reconstructions using L+S and k-t FASTER methods. 
In this experiment, the parameters for each algorithm were tuned for a training sequence, and the results are evaluated using the same parameters for an unseen data sequence. For PEAR, $\lambda$ was examined in the range between 0.25 and 2, and was selected as $\lambda=1.75$ experimentally. For L+S, $\lambda_{1}$ was examined in the range between 0.25 and 1.75, and $\lambda_{2}$ was examined in the range between 0.13 and 1.75. These parameters were selected as $\lambda_1=1.25$ and $\lambda_2=0.25$ experimentally (the values for $\lambda,\lambda_{1,2}$ are provided after normalization with respect to the standard deviation of the data).

\begin{figure}
\begin{flushleft}
\hspace{90mm} Reconstruction results (R=6.66)
\vspace{-2mm}
\end{flushleft}
\includegraphics[width=530pt]{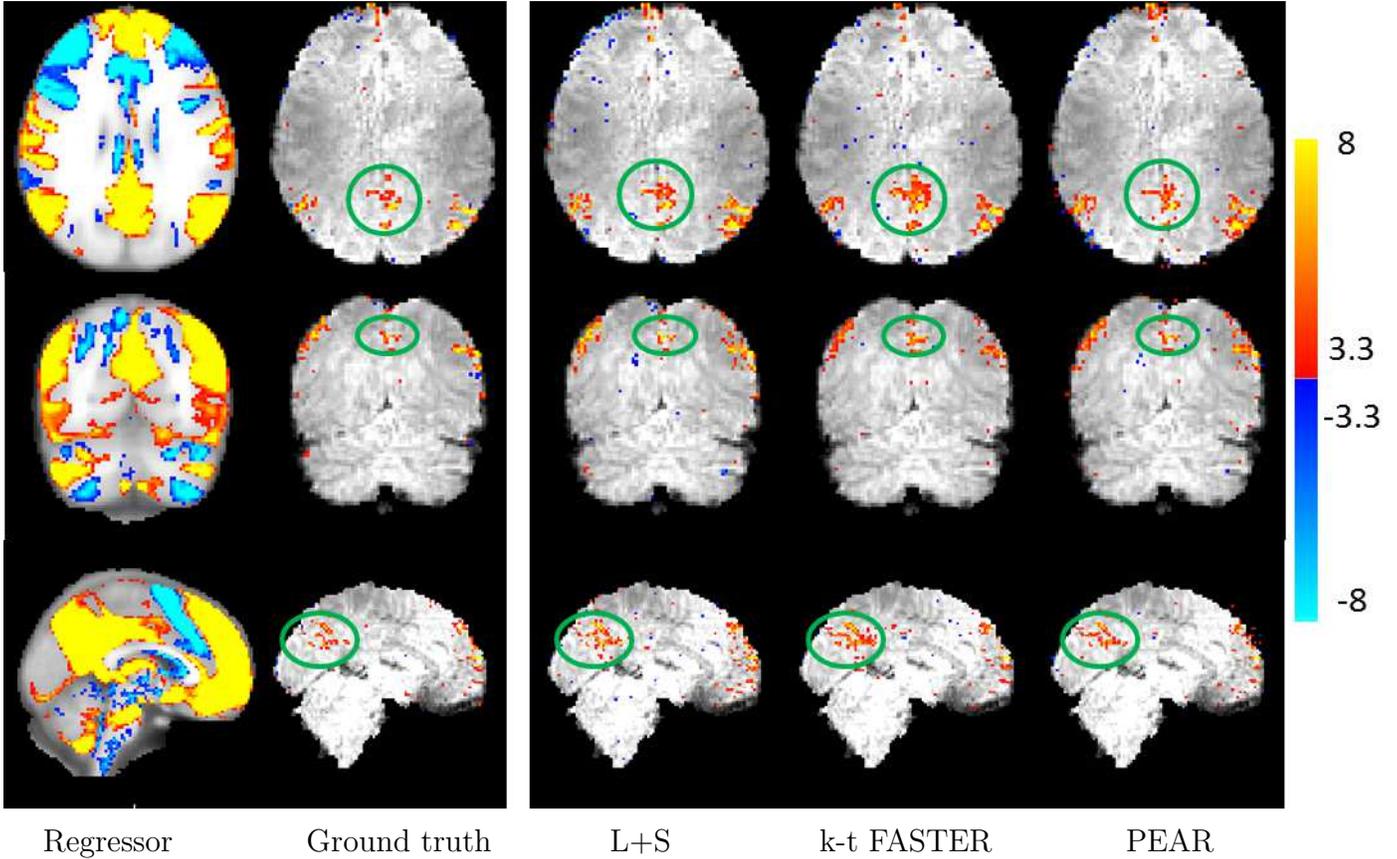}
\begin{flushleft}
\vspace{-3mm}
\hspace{6mm} Regressor \hspace{15mm}  Ground truth \hspace{13mm} L+S \hspace{17mm} k-t FASTER \hspace{15mm} PEAR
\end{flushleft}
\caption{Experiment 3: Retrospective sampling of realistic fMRI dataset.  Left: the regressor used for dual regression overlaid on the MNI template, and the dual regression results of the ground truth. Right: Dual regression results of L+S, k-t FASTER and PEAR obtained at undersampling ratio of 15\% (R=6.66). All maps are thresholded at $|Z|>3.3$ and with color scale mapped between $3.3<|Z|<8$. The green ellipses illustrate regions where the activation pattern detected by PEAR is the most similar one to the ground truth.}
\label{fig7}
\end{figure}

\begin{figure}
\begin{flushleft}
 \hspace{55mm} Reconstruction results (R=10)
\vspace{-2mm}
\end{flushleft}
\includegraphics[width=350pt]{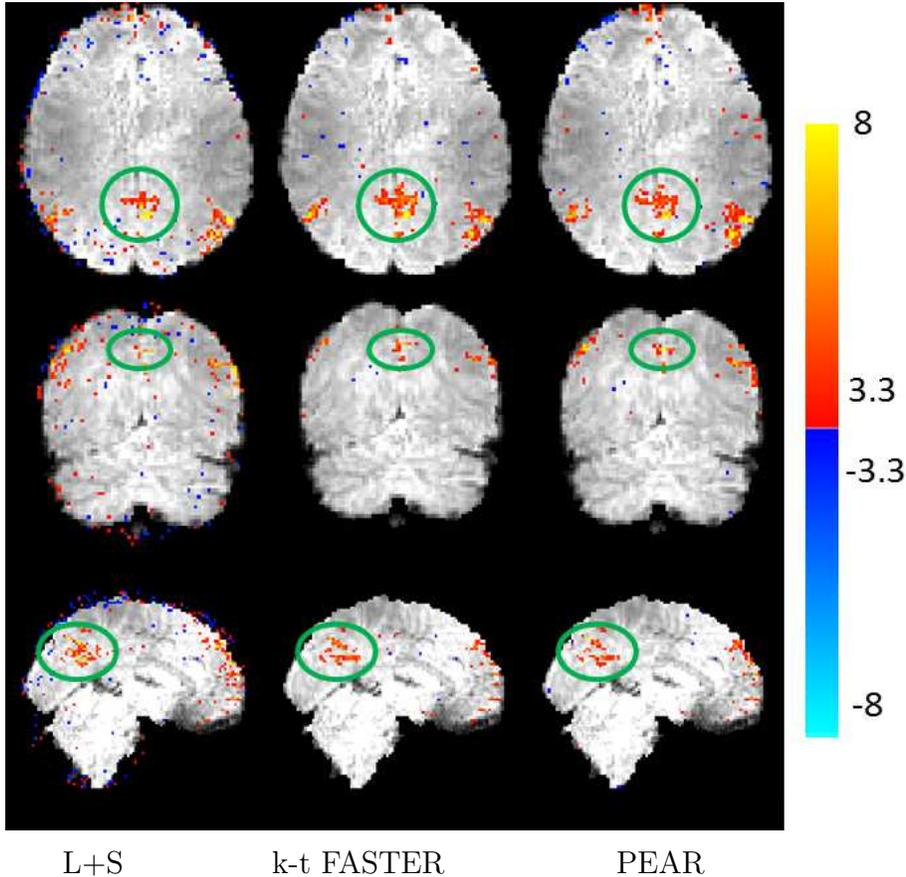}
\begin{flushleft}
\vspace{-3mm}
\hspace{40mm} L+S \hspace{17mm} k-t FASTER \hspace{20mm} PEAR
\end{flushleft}
\caption{Experiment 3: Retrospective sampling of realistic fMRI dataset. Dual regression results of L+S, k-t FASTER and PEAR obtained at undersampling ratio of 10\% (R=10) . All maps are thresholded at $|Z|>3.3$ and with color scale mapped between $3.3<|Z|<8$. The green ellipses illustrate regions where the activation pattern detected by PEAR is the most similar one to the ground truth (shown in Fig. \ref{fig7}). It can be seen that L+S provides very noisy results at this undersampling ratio.
}\label{fig7a}
\end{figure}

Figure \ref{fig7} shows the Default Mode Network (DMN) regressor used for dual regression overlaid on the MNI atlas, as well as  the z-stat dual regression outputs of the ground truth, kt-FASTER, L+S and PEAR for R=6.66. All maps are thresholded at $|Z|>3.3$ and with color scale mapped between $3.3<|Z|<8$.  The results for R=10, are shown in Fig. \ref{fig7a}. The green ellipses in the images show regions where PEAR's activation pattern is the most similar one to the ground truth. It can also be seen that while both k-t FASTER and PEAR provide reliable results for both R=6.66 and R=10, L+S is does not provide satisfactory results at the higher acceleration ratio of R=10.

\begin{figure}
\includegraphics[width=250pt]{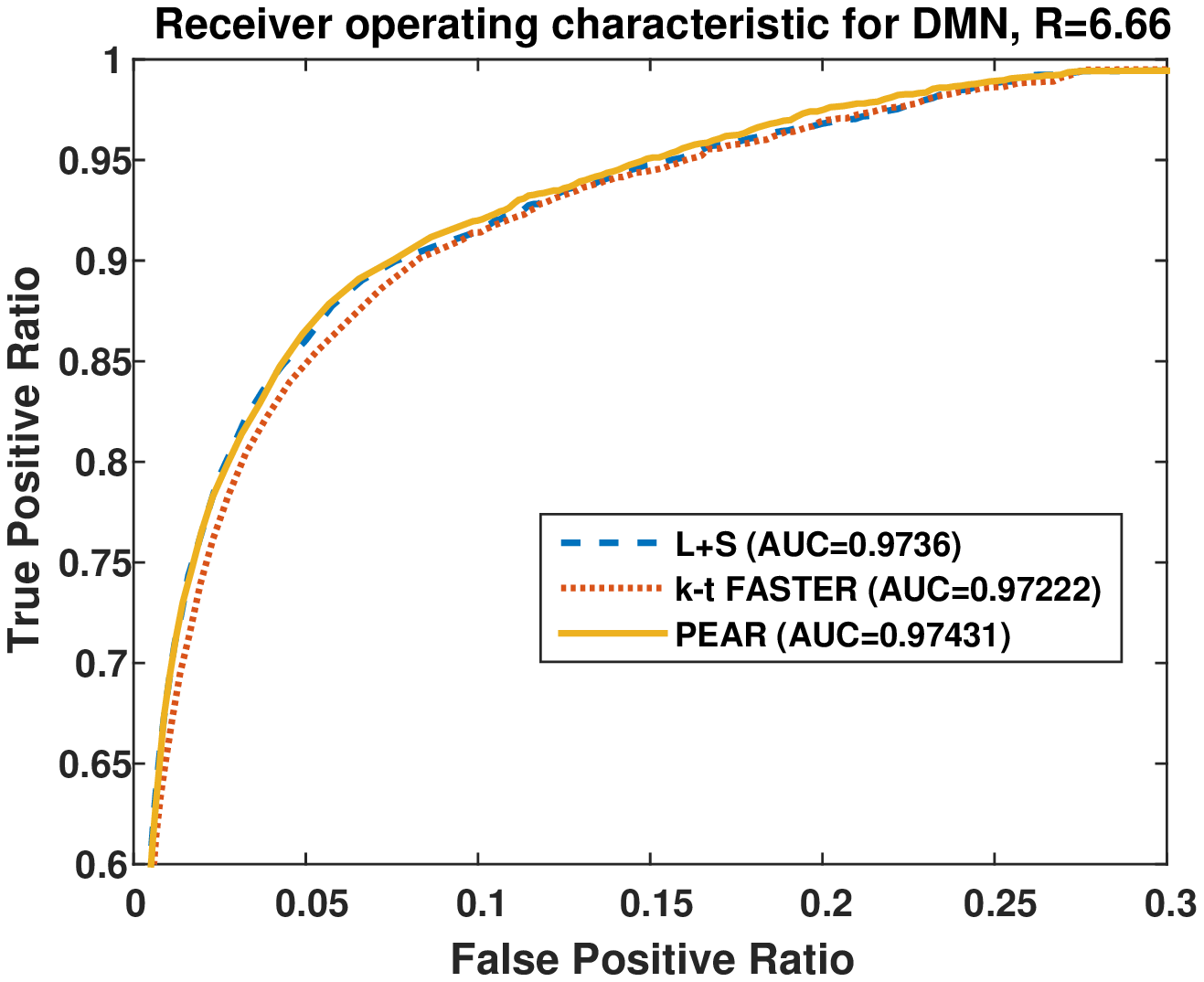}
\includegraphics[width=250pt]{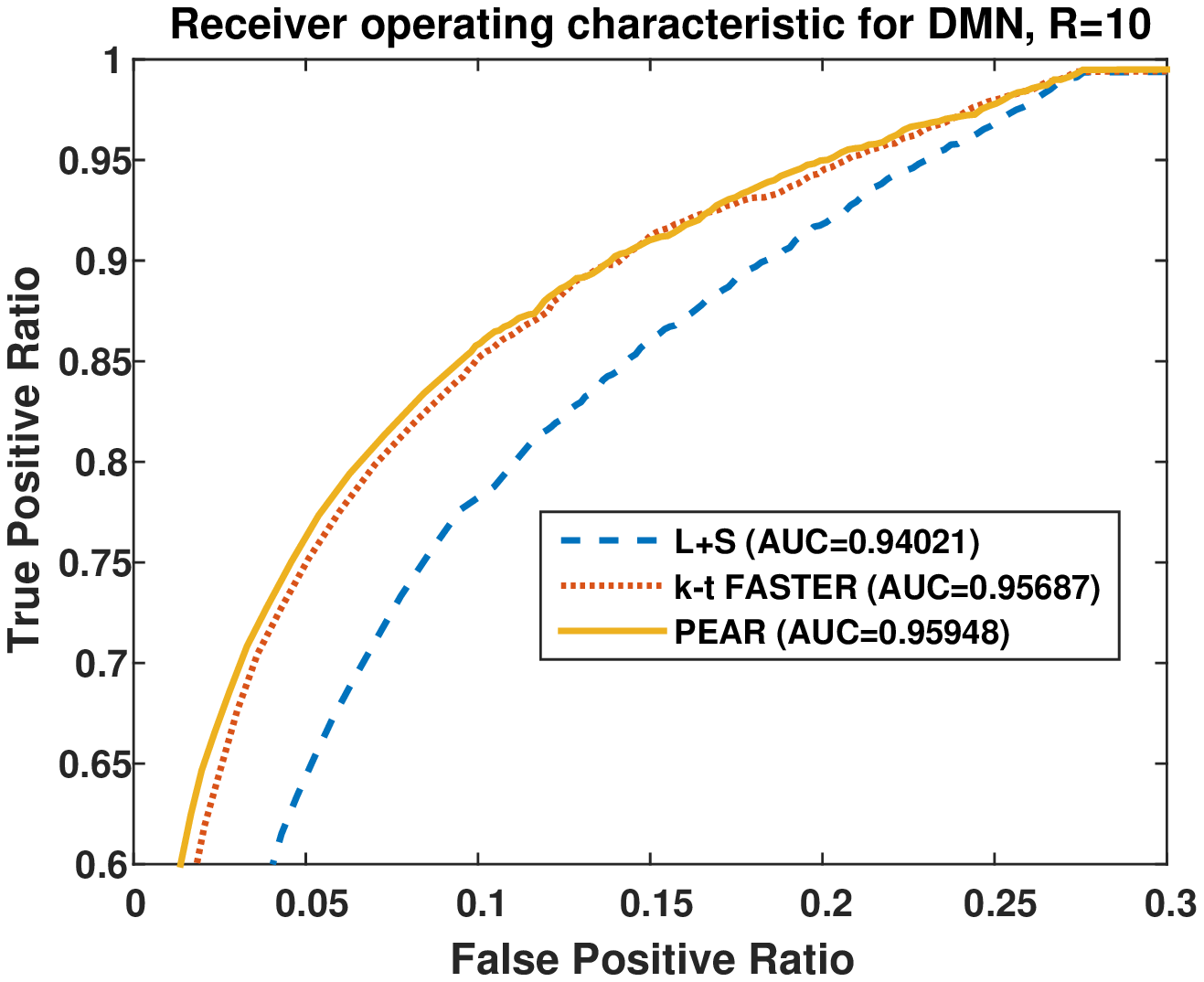}

\caption{Receiver operating characteristic (ROC) curve for experiment 3. Performance of L+S, kt-FASTER and PEAR are shown. The numbers in brackets indicate the area under the curve (AUC) for each method. The ROC results are generated for the ground truth map thresholded at $|Z|>3.3$ serving as a reference.  It can be seen that PEAR provides the highest AUC for DMN, for both R=6.66 and R=10.
}
\label{fig8}
\end{figure}

To provide measure for comparison between the methods, we computed the receiver operating characteristic (ROC) curves. This curve shows the performance of each method when compared to the ground truth (in terms of true positive ratio (TPR) against false positive ratio (FPR)) as the discrimination threshold varies. As a reference, we used the ground truth DMN z-stat map shown in Fig. \ref{fig7} (thresholded at $|Z|>3.3$). For the generation of ROC we computed the TPR and FPR for each DMN z-stat map for each method, as the threshold Z value ranges between 0 and 10. A common scalar measure for the performance of the algorithm is the area under the curve (often referred to as AUC). Figure \ref{fig8} shows the ROC curves for kt-FASTER, L+S and PEAR, including the AUC computed for each curve, for R=6.66 and R=10. It can be seen that PEAR provides the most convex shape with the highest AUC in both cases. In addition, the degraded performance of L+S for R=10 can clearly be seen by examining its ROC curve for R=10. To check the validity of this result for different RSN maps, we computed the AUC for all the 15 canonical RSN maps used in our dual regression process. The summary of the results is shown in Table \ref{table1}, where the method that provides the highest AUC for each RSN in marked in bold (for R=6.66 and R=10 separately). We see that while there are cases in which k-t FASTER or L+S outperform PEAR for some of the maps, PEAR provides the best AUC for the majority of the maps for both R=6.66 and R=10. The results of the training dataset for R=6.66 are also shown in the table, for completeness.

\begin{figure}
\includegraphics[width=200pt]{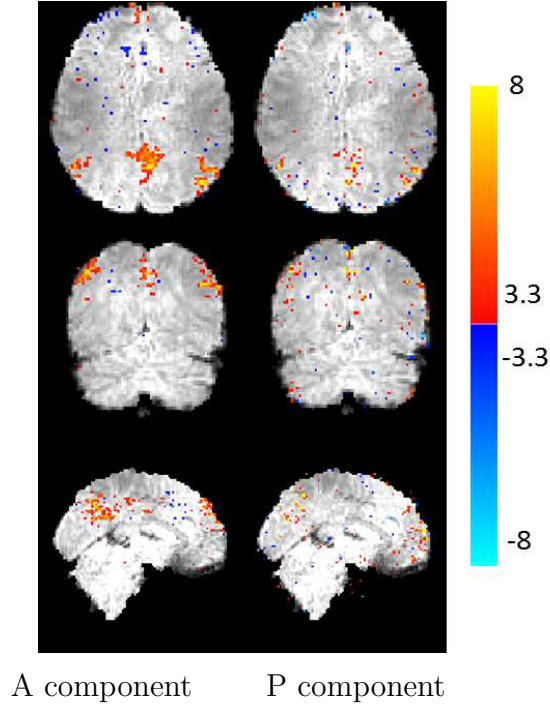}
\begin{flushleft}
\vspace{-3mm}
\hspace{55mm} A component \hspace{7mm} P component
\end{flushleft}
\caption{Z-stat maps of A and P components of PEAR for R=6.66. Maps are thresholded at $|Z|>3.3$ and with color scale mapped between $3.3<|Z|<8$. It can clearly be seen that both components contain functional information and present activation regions in locations that correspond to similar activation regions in the ground truth.}
\label{fig11}
\end{figure}

\begin{figure}
\includegraphics[width=520pt, trim={6.5cm 0cm 4.5cm 0cm},clip=true]{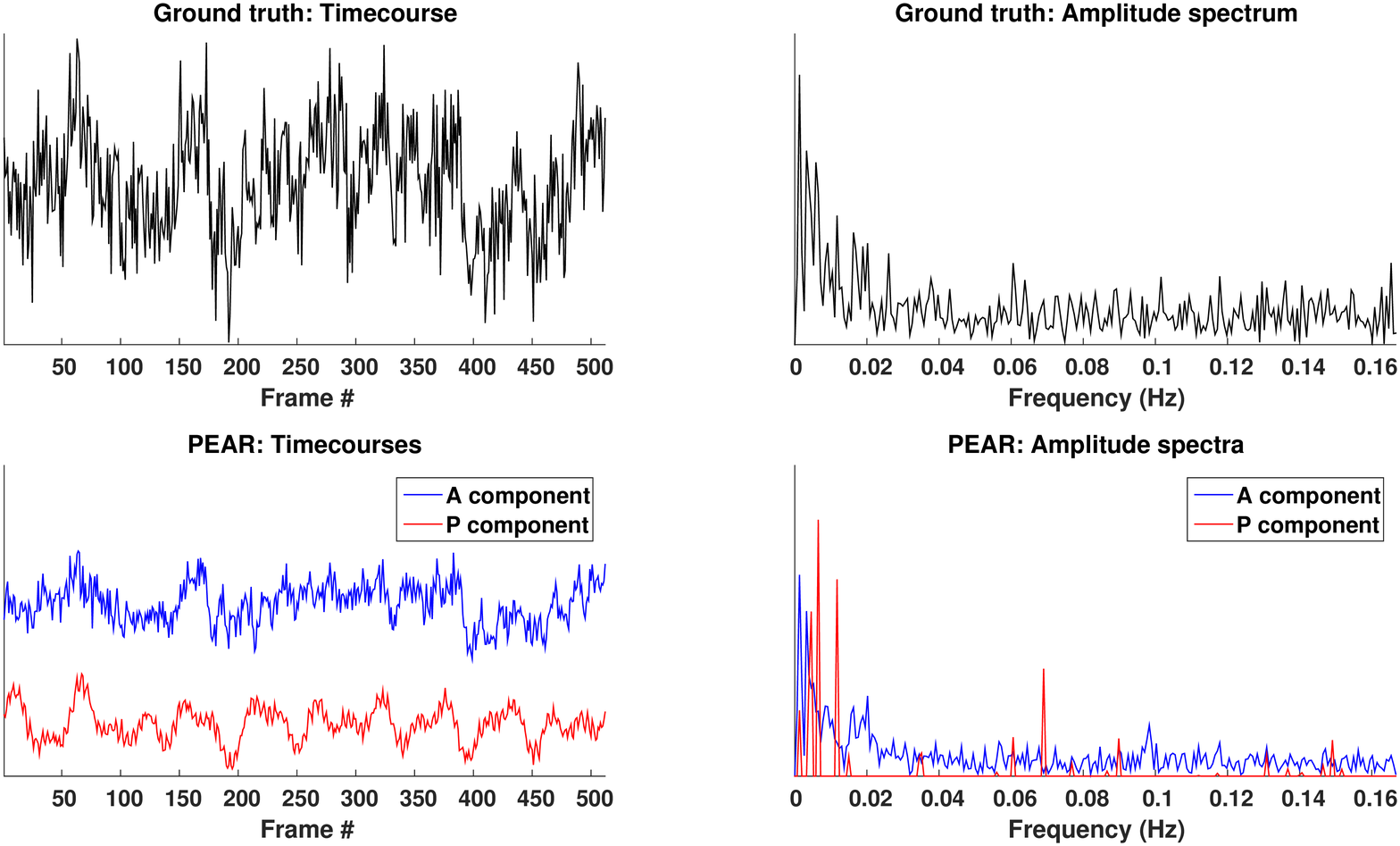}\\
\begin{minipage}[c]{0.3\textwidth}
\includegraphics[width=110pt]{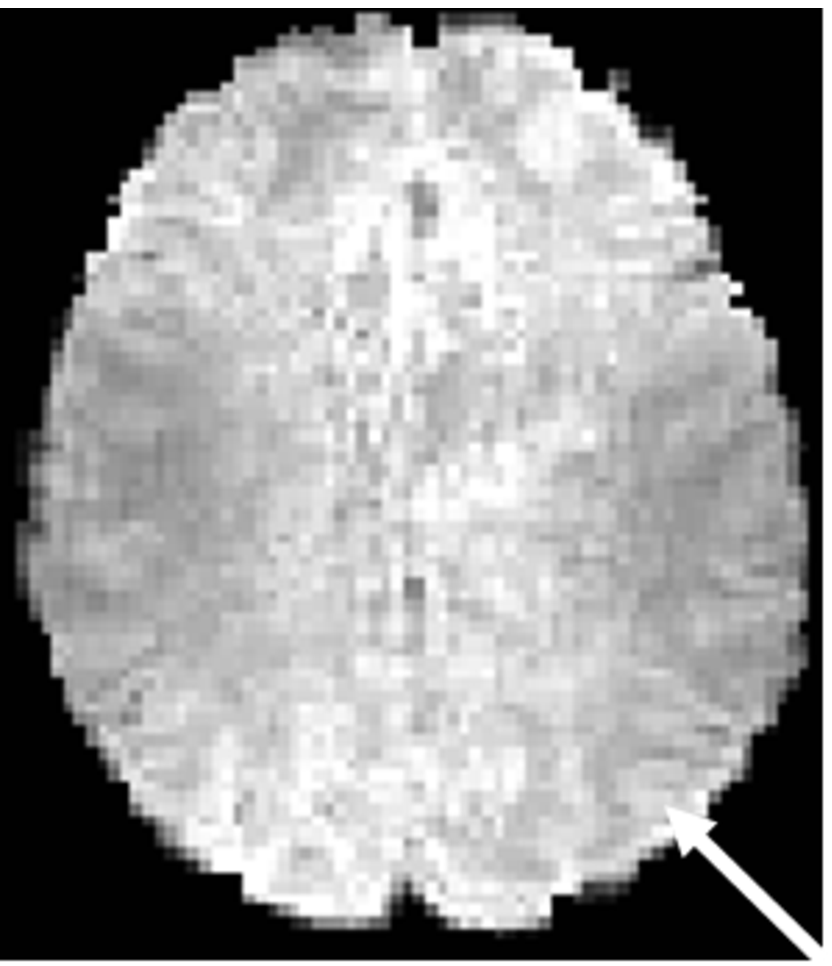}
  \end{minipage}
\begin{minipage}[c]{0.65\textwidth}
\caption{Top: Timecourses and amplitude spectra of a pixel resulted with $|Z|>4.5$ for DMN z-stat map. Timecourses and amplitude spectra of ground truth (top row) and A and P components of PEAR result (for R=6.66) (second row) are shown. Values are shown in arbitrary units for better view. The spatial location of the selected pixel on an axial slice is also shown (left). It can be seen that the A component contains a wide range of frequencies, with some strong peaks in the spectrum for frequencies that have strong total energy. The P component contains a limited number of temporal frequencies and captures the periodicity that is not captured in the fixed-rank component, A. This separation shows that a selection of a fixed, moderate rank for the A component, in addition to a demand for a limited number of temporal frequencies for the P component leads to the desired separation, which results in improved z-stat results shown earlier.}
\label{fig9}
  \end{minipage}
\end{figure}

Finally, we examined the separation of PEAR into into A and P components, in terms of both time courses and spatial z-stat maps (for R=6.66). For this purpose, we first performed dual regression analysis for the A component and for the P component separately, to generate a z-stat map for each. Those maps, thresholded at $|Z|>3.3$ and with color scale mapped between $3.3<|Z|<8$, are shown in Fig. \ref{fig11} and demonstrate that both A and P components contain functional activity. Then, we arbitrary selected a single pixel that exhibited high correspondence with DMN for both A and P (z-stat value of $|Z|>4.5$). Figure \ref{fig9} shows the timecourses and amplitude spectra of the A and P components, for the selected pixel, where the spatial location of the pixel is shown at the bottom of the Figure. The timecourses and amplitude spectra show that the A component contains a wide range of frequencies, with some strong peaks in the spectrum for frequencies that have strong total energy. The P component contains a limited number of temporal frequencies and captures the low-energy periodicity that is not captured in A. This separation shows that a selection of a fixed, moderate rank for the A component, in addition to a demand for a limited number of temporal frequencies for the P component leads to the desired separation, which results in improved z-stat results shown earlier.

\begin{table}
\caption{Area under ROC for 15 RSN maps. The bold values indicate the method with the highest value in each line, for R=6.66 and R=10 separately. It can be seen that PEAR provides the best performance for most of the maps, for both R=6.66 and R=10. Training dataset results for R=6.66 are also shown for completeness.}
  \begin{tabular}{ | m{1.2cm} | m{1.55cm} | m{1.9cm}| m{1.55cm} || m{1.55cm} | m{1.9cm}| m{1.55cm} ||m{1.55cm}|m{1.9cm}|m{1.55cm}|}
    \hline
    \multirow{3}{*}{\vtop{\hbox{\strut RSN}\hbox{\strut Map \#}}}& \multicolumn{3}{c||}{Training dataset}&\multicolumn{6}{c|}{Testing dataset}\\\cline{2-10}
    &\multicolumn{3}{c||}{R=6.66}&\multicolumn{3}{c||}{R=6.66}&\multicolumn{3}{c|}{R=10}\\ \cline{2-10} 
& L+S & kt-FASTER & PEAR & L+S & kt-FASTER & PEAR &L+S & kt-FASTER & PEAR \\ \hline \hline
1& 0.96277 &    0.96577  &  {\bf  0.96781 }& 0.96877 &                  0.96883      &            {\bf 0.97277}      &             0.93883     &              0.95361      &            {\bf 0.95615}\\ \hline
2& 0.97591  &    0.9769  &  {\bf  0.97924 }& 0.9736  &                 0.97222        &       {\bf    0.97431}       &            0.94021      &             0.95687       &          {\bf  0.95948}\\ \hline
3& 0.97874 &    0.97789  &  {\bf  0.98244 }& 0.97933 &                  0.97794       &          {\bf  0.98208 }     &             0.96465     &              0.96612      &           {\bf  0.96712}\\ \hline
4& 0.9754  &   0.97446  &  {\bf  0.97754 }& 0.97658 &                  0.97452       &           {\bf 0.97728}      &             0.92735     &              {\bf 0.95559}      &             0.95503\\ \hline
5& 0.97529  &   0.97461   &{\bf   0.97708 }& 0.97526 &                  0.97688       &           {\bf 0.97755}      &             0.94546     &          {\bf    0.95622}      &             0.94724\\ \hline
6& 0.97462  &   0.97433   &    {\bf 0.976} & {\bf 0.97239} &                  0.97238       &            0.97091      &           {\bf  0.94525}     &              0.94523      &             0.93694\\ \hline
7& 0.96876 &    0.97043  &   {\bf 0.97082} &  0.96444 &                  0.96823      &        {\bf     0.96846}      &             0.92203     &               0.9363      &           {\bf  0.93836}\\ \hline
8& 0.97306  &   0.97881   & {\bf  0.97935} &  0.9784  &                 0.97938       &        {\bf    0.98079}       &             0.9628      &             0.96357       &           {\bf 0.96555}\\ \hline
9& 0.95426   &  0.95392 &   {\bf  0.95535 }& 0.95299 &                  0.95289     &             {\bf 0.95464}      &             0.91514     &          {\bf    0.93129}      &             0.92877\\ \hline
10&0.97721  &   0.97639    &{\bf  0.97778} & 0.97244     &            {\bf  0.97562}      &             0.97369     &              0.94051    &               0.94603     &              {\bf 0.9479}\\ \hline
11& 0.96271  &  {\bf  0.97008}  &   0.96684& 0.97813     &             {\bf 0.97901}      &             0.97835     &             {\bf 0.95839}    &               0.95413     &              0.95635\\ \hline
12& 0.97276  &  {\bf  0.97506 }   &  0.9747 & 0.96785     &           {\bf    0.9698}      &              0.9693     &              0.90773    &               0.94283     &          {\bf    0.94328}\\ \hline
13&0.97282  &   0.97967  &   {\bf 0.98012} & 0.96625     &              0.97007      &             {\bf 0.97086}     &              0.92664    &               {\bf 0.93289}     &              0.93045\\ \hline
14&0.97   &  0.97148     &{\bf  0.9722 }& 0.96661     &             {\bf 0.96741}      &             0.96622     &              0.92057    &               {\bf 0.94641}     &              0.93858\\ \hline
15& 0.97133   &  0.97555  & {\bf   0.97695 }& 0.97087     &              0.97282      &            {\bf 0.97465}     &              0.95221    &               0.94615     &              {\bf 0.95443}\\ \hline
  \end{tabular}
  \label{table1}
\end{table}

\section{Discussion}
\label{sec:discussion}
\subsection{Relation to previous works}
As described in the Introduction, a few methods that have been published recently also focus on separation of fMRI into two components \cite{lam2013accelerated,singh2015under,aggarwal2017optshrink,otazo2015Lowrankplus}. The main differences between our work and these prior works lie in the methodology and experiments. First, we enforce both components to contain functional information explicitly, by solving an unconstrained minimization problem that enforces the A component to have a fixed moderate rank, using a TSVD based solution. This is in contrast to L+S-based methods that practically enforce all the functional information to be contained in a single component, required to be sparse in some transform domain. As a result, the L+S solution might be sub-optimal in reconstruction of signals that are neither periodic, nor strong enough to be captured in the low-rank component. 

Second, the experimental part of this paper presents a thorough validation of the suggested approach (compared to other existing methods), based on realistic nonuniform undersamping, and examines the spatial activation of resting state networks with broad spectrum characteristics. This analysis, which also includes the examination of the functional information that is contained in each of the components, is expanded compared to previous papers that deal with separation of fMRI into components; in particular, some of them show task-based MRI where the design is periodic or basic RSN analysis.


\subsection{Error measures and reproducibility}
It is important to note that in the case of fMRI, conventional measures between the reconstructed datasets (e.g. MSE or correlations) may be misleading, as lower MSE does not necessarily mean that low variance functional information is preserved. Therefore, evaluation of results in this work is based the z-stat analysis of activation maps, which is the common tool used today for resting state fMRI analysis. 

In addition, while prospective undersampling is possible (and has been carried out in our previous works \cite{chiew2015k}), its evaluation has to be performed against connectivity maps taken from the literature or group-averaged RSN spatial maps, leading to uncertainty in ensuring that subject-specific detail is retained. Therefore, we focus on retrospectve undersampling, which allows accurate comparison against a well defined ground truth.




\subsection{Limitations}
This work focuses on resting state fMRI, which is a branch in fMRI research that deals with mapping brain connectivity based on an fMRI experiment that does not involve stimulation. In contrast, task-based fMRI involves known manipulations of brain activity (a.k.a task-based fMRI), although some of the properties of task-based fMRI data are very similar to resting fMRI (i.e. the low variance signal based on the BOLD effect). Since many other reconstruction techniques place strong assumptions on brain activity, e.g. that it is periodic, highly reproducible or smooth in time, PEAR does not place these assumptions and therefore is expected to provide reliable results also for task-base fMRI. However, the analysis of task-based fMRI with PEAR is reserved for future research.

In addition, the reconstruction time of PEAR for a single realistic 2D axial slice in the dimensions described in our experimental results ($109\times 91$ with 512 time points) is approximately 15 minutes using MATLAB running on a single machine with 3.2GHz CPU. To allow recovery of a multi-slice image, we used cluster-based computing that processed all 91 slices in parallel and provided a 3D volume approximately at the same running time. While the computation time and the need for cluster-based computing are drawbacks of all iterative methods used to reconstruct fMRI sequences examined in this paper (k-t FASTER and L+S), it is performed offline, while the subject is no longer in the scanner and does not require expensive MRI resources. In addition, we are examining approches to accelerate the reconstruction process, mainly by improving the NUFFT, using methods with lower computational complexity\cite{kiperwas2017spurs} and implementing algorithms in a real-time programming environment.

\section{Conclusions}
This paper presents PEAR, an under-sampled fMRI reconstruction approach based on separating the fMRI signal to periodic and fixed-rank components.  The higher accelaration ratio offered by PEAR results in reconstruction with higher fidelity than when using a fixed-rank based model or a conventional L+S algorithm. We have shown that splitting the functional information between the A and P components, by solving a constrained problem that enforces a fixed, moderate rank for the A component, leads to better modeling for fMRI, due to the unique nature of the fMRI signal. Future work will focus on extending this work to task-based fMRI using both retrospective and prospective sampling.

\label{sec:conclusions}

\section*{Acknowledgements} This work was supported by the Israeli Ministry of Science, by the ISF I-CORE joint research center of the Technion and the Weizmann Institute, Israel, by the European Union's Horizon 2020 research and innovation programme under grant agreement No. 646804-ERC-COG-BNYQ, by the British Technion Society and the Coleman-Cohen Fellowship for post-doctoral studies in the UK, and by the EPSRC and the Wellcome Trust.

\noindent The authors have no relevant conflicts of interest to disclose.

\appendix
\section{Solution of (\ref{eq22}) using ISTA}
\label{sec:appendix}
Let $\mat{Q}=\mat{F}_t\{\mat{P}\}$. Using the fact that $\mat{F}_t$ is unitary, (\ref{eq22}) can be written as:
\begin{equation}
\mat{P}_n=\mat{F}_t^H\{D(\mat{A}_n,\mat{Q})\}
\label{eqa1}
\end{equation}
\noindent where
\begin{equation}
D(\mat{A}_n,\mat{Q})=\underset{ \mat{Q}\in R^{N \times K}}{\text{arg min}}  \frac{1}{2}\|\vec{y}-\mat{E}\{\vec{A}_n+\mat{F}_t^H\{\mat{Q}\}\}\|_2^2+\lambda\|\mat{Q}\|_1.
\label{eq7}
\end{equation}
An iterative solution to (\ref{eq7}) can be obtained using the iterative shrinkage-thresholding algorithm (ISTA)\cite{daubechies2004iterative,combettes2005signal,beck2009fast} ,  whose general step is:
\begin{equation}
\mat{Q}_{k+1}=\Lambda_{\lambda}(\mat{Q}_{k}-\alpha\mat{F}_t\{\mat{E}^H\{\mat{E}\{\mat{A}_n+\mat{F}_t^H\{\mat{Q}_k\}\}-\mat{y}\}\}) \label{eq391}
\end{equation}
\noindent Here $\Lambda_{\lambda}$ is the soft-thresholding operator with parameter $\lambda$, $\alpha$ is the step size and $\mat{Q}_0=\mat{P}_{n-1}$. Using the fact that $\mat{P}_n=\mat{F}_t^H\{\mat{Q}_K\}$, setting $K=1$ and using (\ref{eqa1}) we get that the general step for the solution of (\ref{eq22}) is:
\begin{equation}
\mat{P}_{n}=\mat{F}_t^H\{\Lambda_\lambda(\mat{F}_t\{\mat{P}_{n-1}\}-\alpha\mat{F}_t\{\mat{E}^H\{\mat{E}\{\mat{A}_n+\mat{P}_{n-1}\}-\mat{y}\}\})=\mat{F}_t^H\{\Lambda_\lambda(\mat{F}_t\{\mat{P}_{n-1}-\alpha\mat{E}^H\{\mat{E}\{\mat{A}_{n-1}+\mat{P}_{n-1}\}-\vec{y}\}    \})\},
\label{eqa2}
\end{equation}
\noindent where the last equality is (\ref{eq25}) in the paper.


\bibliography{PEAR_journal}

\end{document}